\def\lesssim{{_ <\atop{^\sim}}}

\def\ap3m{AP$^3$M}
\def\LCDM{$\Lambda$CDM}

\def\hkpc{$h^{-1}{\ }{\rm kpc}$}
\def\hMpc{$h^{-1}{\ }{\rm Mpc}$}
\def\hMsun{$h^{-1}{\ }{\rm M_{\odot}}$}
\def\kms{${\rm{\ }km{\ }s^{-1}}$}
\def\nbody{$N$-body}
\def\llink{$b_{\rm link}$}
\def\c15{$c_{\rm 1/5}$}

\def\Dsat{$D_{\rm sat}$}
\def\rmhf{$r_{\tt MHF }$}
\def\rvir{$r_{\rm vir}$}
\def\Rvir{$R_{\rm vir}$}
\def\Mvir{$M_{\rm vir}$}

\def\rtidal{$r_{\rm tidal}$}
\def\zform{$z_{\rm form}$}
\def\rsmooth{$r_{\rm smooth}$}

\newcommand{\Table}[1]{Table~\ref{#1}}
\newcommand{\Sec}[1]{Section~\ref{#1}}
\newcommand{\Eq}[1]{Equation~(\ref{#1})}
\newcommand{\Fig}[1]{Figure~\ref{#1}}
\newcommand{\mlapm}{\texttt{MLAPM}}
\newcommand{\mhf}{\texttt{MHF}}
\newcommand{\mht}{\texttt{MHT}}
\newcommand{\finder}{\texttt{Finder}}
\newcommand{\tracker}{\texttt{Tracker}}

\newcommand{\GKGD}{\textsf{GKGD}\textrm{II}}

\def\ea{et~al.~}                            % \ea      =  et al.
\def\lesssim{\mathrel{\hbox{\rlap{\hbox{\lower4pt\hbox{$\sim$}}}\hbox{$<$}}}}
\def\gtrsim{\mathrel{\hbox{\rlap{\hbox{\lower4pt\hbox{$\sim$}}}\hbox{$>$}}}}

\newcommand{\CP}[3]    {\mbox{Computers in Physics~\textbf{#1},~#2~(#3)}}

\newcommand{\ApJ}[3]    {\mbox{ApJ~\textbf{#1},~#2~(#3)}}
\newcommand{\ApJS}[3]   {\mbox{ApJ~Suppl.~\textbf{#1},~#2~(#3)}}
\newcommand{\ApJL}[3]   {\mbox{ApJ~Lett.~\textbf{#1},~#2~(#3)}}
\newcommand{\ARAA}[3]   {\mbox{Ann.~Rev.~A~\&~A~\textbf{#1},~#2~(#3)}}
\newcommand{\AJ}[3]     {\mbox{Astron.~J.~\textbf{#1},~#2~(#3)}}
\newcommand{\MNRAS}[3]  {\mbox{MNRAS~\textbf{#1},~#2~(#3)}}
\newcommand{\Nature}[3] {\mbox{Nature~\textbf{#1},~#2~(#3)}}

\newcommand{\astroph}[1]{\mbox{\texttt{astro-ph/#1}}}

\documentstyle[epsfig]{mn}

\begin{document}
%%%%%%%%%%%%%%%%%%%%%%%%%%%%%%%%%%%%%%%%%%%%%%%%%%%%%%%%%%%%%%%%%%%%%%%%
%                               TITLE                                  %
%%%%%%%%%%%%%%%%%%%%%%%%%%%%%%%%%%%%%%%%%%%%%%%%%%%%%%%%%%%%%%%%%%%%%%%%
\title[Identifying dark matter substructure]{The evolution substructure I: a new identification method}

\author[Gill S.P.D., Knebe A. \& Gibson B.K.]
       {Stuart P.~D. Gill, Alexander Knebe, Brad K. Gibson\\        
       {Centre for Astrophysics \& Supercomputing,
        Swinburne University, Mail \#31, P.O. Box 218,
        Hawthorn, Victoria, 3122, Australia}\\}

\date{Received ...; accepted ...}

\maketitle

%%%%%%%%%%%%%%%%%%%%%%%%%%%%%%%%%%%%%%%%%%%%%%%%%%%%%%%%%%%%%%%%%%%%%%%%
%                             ABSTRACT                                 %
%%%%%%%%%%%%%%%%%%%%%%%%%%%%%%%%%%%%%%%%%%%%%%%%%%%%%%%%%%%%%%%%%%%%%%%%
\begin{abstract}
We describe our new "\mlapm-halo-finder" (\mhf) which is based on the
adaptive grid structure of the \nbody\ code \mlapm. We then extend the
\mhf\ code in order to track the orbital evolution of
gravitationally bound objects through any given cosmological
\nbody-simulation - our so-called "\mlapm-halo-tracker" (\mht). The
mode of operation of \mht\ is demonstrated using a series of eight
high-resolution \nbody\ simulations of galaxy clusters.  Each of these
halos hosts more than one million particles within their virial radii
\rvir. We use \mht\ as well as \mhf\ to follow the temporal evolution
of hundreds of individual satellites, and show that the radial
distribution of these substructure satellites follows a ``universal''
radial distribution irrespective of the host halo's environment and
formation history. This in fact might pose another problem for
simulations of CDM structure formation as there are recent findings by
Taylor~\ea (2003) that the Milky Way satellites are found preferentially
closer to the galactic centre and simulations underestimate the amount
of central substructure, respectively. Further, this universal
substructure profile is anti-biased with respect to the underlying
dark matter profile. Both the halo finder
\mhf\ and the halo tracker \mht\ will become part of the open source
\mlapm\ distribution.
\end{abstract}


%%%%%%%%%%%%%%%%%%%%%%%%%%%%%%%%%%%%%%%%%%%%%%%%%%%%%%%%%%%%%%%%%%%%%%%%
%                           INTRODUCTION                               %
%%%%%%%%%%%%%%%%%%%%%%%%%%%%%%%%%%%%%%%%%%%%%%%%%%%%%%%%%%%%%%%%%%%%%%%%
\section{Introduction}

Over the last 30 years great progress has been made in the development of
\nbody\ codes that model the distribution of dissipationless dark matter.
Algorithms have advanced considerably since the first
$N^2$ particle-particle codes (Aarseth 1963; Peebles 1970; Groth \ea
1977); we have seen the development of the tree-based gravity solvers
(Barnes~\& Hut 1986), mesh-based solvers (Klypin~\& Shandarin 1983), then
the two combined (Efstathiou \ea 1985) and multiple strands of adaptive
and deforming grid codes (Villumsen 1989; Suisalu~\& Saar 1995; Kravtsov,
Klypin~\& Khokhlov 1997; Bryan~\& Norman 1998; Knebe, Green \& Binney
2001). While they all push the limits of efficiency in computational
resources, each code has its individual advantages and limitations. The
result of such research has been highly reliable, cost effective codes.
However, producing the data is only one step in the process; the ensembles
of millions of (dissipationless) dark matter particles generated still
require interpreting and then comparison to the real Universe. This 
necessitates access to
analysis tools to map the phase-space which is being sampled by
the particles onto ``real'' objects in the Universe; traditionally this has
been accomplished through the use of ``halo finders''. Halo finders mine
\nbody\ data to find locally over-dense gravitationally bound systems,
which are then attributed to the dark halos we currently believe surround
galaxies. Such tools have lead to critical insights into our understanding
of the origin and evolution of structure and galaxies. To take advantage
of sophisticated \nbody\ codes and to optimise their predictive power one
needs an equally sophisticated halo finder.

Over the years, halo-finding algorithms have paralleled the development of
their partner \nbody\ codes. We briefly outline the major halo finders
currently in use:

The Friends-of-Friends (FOF) (Davis \ea 1985; Frenk \ea 1988) algorithm
uses spatial information to locate halos. Specifying a linking length
\llink\ the finder links all pairs of particles with separation equal to
or less than \llink\ and calls these pairs ``friends''. Halos are defined by
groups of friends (friends-of-friends) that have at least one of these
friendship connections. Two such advantages of this algorithm are its ease
of interpretation and its avoidance of assumption concerning the halo shape. 
The greatest disadvantage is its simple choice of linking length which can
lead to a connection of two separate objects via so-called linking
``bridges''. Moreover, as structure formation is hierarchical, each halo
contains substructure and thus the need for different linking lengths to
identify ``halos-within-halos''.  There have been many variants to this
scheme which attempt to overcome some of these limitations (Suto, Cen \&
Ostriker 1992; Suginohara \& Suto 1992; van Kampen 1995; Okamoto \& Habe
1999; Klypin \ea 1999).

DENMAX (Bertschinger \& Gelb 1991; Gelb \& Bertschinger 1994a) and SKID
(Weinberg, Hernquist \& Katz 1997) are similar methods in that they both
calculate a density field from the particle distribution, then gradually
move the particles in the direction of the local density gradient
ending with small groups of particles around each local density maximum.
The FOF method is then used to associate these small groups with
individual halos. A further check is employed to ensure that the grouped
particles are gravitationally bound.  The two methods differ through
their calculation of the density field. DENMAX uses a grid while SKID
applies an adaptive smoothing kernel similar to that employed
in Smoothed Particle Hydrodynamics techniques
(Lucy 1977; Gingold~\& Monaghan 1977; Monaghan 1992). The effectiveness of
these methods is limited by the method used to determine the density field
(G\"otz, Huchra \& Brandenberger 1998).
 
A similar technique to the above is the Bound Density Maxima (BDM) method
(Klypin \& Holtzman 1997; Klypin \ea 1999). In this scheme a smoothed
density is derived by smearing out the particle distribution on a scale
\rsmooth\ of order the force resolution of the \nbody\ code used to
generate the data. Randomly placed ``seed spheres'' with radius \rsmooth\
are then shifted to their local centre-of-mass in an iterative procedure
until convergence is reached. Hence, as with DENMAX and SKID, this process
finds local maxima in the density field. Bullock \ea (2001) further
refined the BDM technique by first generating a set of possible centres,
ranking the particles with respect to their local density and then
implementing modifications which allow for credible identification of
halos-within-halos.  The Bullock~\ea (2001) adaptation to BDM excels at
finding halo substructure.

When one is primarily concerned with distinct halos, all the mentioned
methods perform exceedingly well. All efforts to refine and enhance
those halo finding algorithms are due to the fact that \nbody\ codes
overcame overmerging only recently (Klypin~\ea 1999) and
are capable of finding satellites galaxies within dark matter host
halos.  It is therefore crucial to reliably identify
``halos-within-halos''. In fact, one of the remaining problems for simulations of
CDM structure formation is that 
high-resolution simulations nowadays predict far greater substructure (in
total) than observed (Klypin~\ea 1999; Moore~\ea 1999). Results
from gravitational microlensing suggest that the majority of
substructure which does exist has to be close to the inner regions (Dalal \& Kochanel
2002) which thus far has not been confirmed by such simulations. There
are recent claims that although the overmerging problem has
disappeared in the outer regions of the halo, the inner regions might
still suffer from it (Taylor, Silk~\& Babul 2003). As these latter
semi-analytic models do not suffer from such numerical problems, they
find that such substructure does exist in the inner regions.  The
question though arises as to whether there still remains an overmerging
problem in the simulations or if current halo finding algorithms
actually do break down at those scales. As we will discuss later, it
becomes more difficult to locate peaks in the central region (if at
all present) of the host halo due to a simple lack of contrast.
  
In this paper we present a new method for identifying gravitationally
bound objects in \nbody\ code output that uses the adaptive meshes of
\mlapm\ (Knebe et~al. 2001). This new code excelled at finding
``halos-within-halos'' revealing more substructure in the inner
regions of the host halo.  In its native form, our new algorithm works
naturally ``on-the-fly'', but it has also been constructed with the
flexibility necessary to handle a single temporal output from any
\nbody\ code. Our analysis software will become part of the publicly
available \mlapm\
distribution\footnote{\texttt{http://astronomy.swin.edu.au/MLAPM/}}.
The outline of the paper is as follows. In Section~\ref{Computation}
we introduce the cosmological models used to frame our discussion of
the mode of operation of the new halo finder and tracker. A more
detailed scientific analysis of this data set can be found in Paper~II
of this series (Gill et~al. 2004a; hereafter, \GKGD). In \Sec{mhf} we
introduce the new halo finder ``\mlapm-halo-finder'' (\mhf),
describing its function, advantages, and limitations. \Sec{analmhf}
provides a brief analysis of the satellites found by \mhf. In
Section~\ref{mht} we introduce the ``\mlapm-halo-tracker'' (\mht)
which augments the halo finder by incorporating the ability to track
the temporal evolution of satellites. Analysis of the halos tracked
with \mht\ is described in \Sec{analmht}. We next compare the two
methods with other publicly available halo finding algorithms, such as
FOF and SKID, in Section~\ref{compare}. We conclude with a summary and
our conclusions in \Sec{conclusions}.

This paper is the first in a series of three based upon the
suite of simulations described herein. Paper~II (\GKGD)
investigates the satellite
environments and their dynamical properties, while Paper~III 
(Gill et~al. 2004b) will
investigate the tidal streams and debris from the disrupting satellites.

\begin{table*} \label{HaloDetails}
\caption{Summary of the eight host dark matter halos. Distances are measured
         in \hMpc, velocities in \kms, masses in 10$^{14}$\hMsun, and
         the age in Gyrs.}
\begin{tabular}{ccccccc}\hline
Halo & \Rvir & $V_{\rm circ}^{\rm max}$ & \Mvir  & \zform & age &
$N_{\rm sat}(<r_{\rm vir})$ \\
 
\hline \hline
 \# 1 &  1.34 & 1125 & 2.87 & 1.16 & 8.30 & 158 \\
 \# 2 &  1.06 &  894 & 1.42 & 0.96 & 7.55 &  63 \\
 \# 3 &  1.08 &  875 & 1.48 & 0.87 & 7.16 &  87 \\
 \# 4 &  0.98 &  805 & 1.10 & 0.85 & 7.07 &  57 \\
 \# 5 &  1.35 & 1119 & 2.91 & 0.65 & 6.01 & 175 \\
 \# 6 &  1.05 &  833 & 1.37 & 0.65 & 6.01 &  85 \\
 \# 7 &  1.01 &  800 & 1.21 & 0.43 & 4.52 &  59 \\
 \# 8 &  1.38 & 1041 & 3.08 & 0.30 & 3.42 & 251 \\
\hline
\end{tabular}
\end{table*}

%%%%%%%%%%%%%%%%%%%%%%%%%%%%%%%%%%%%%%%%%%%%%%%%%%%%%%%%%%%%%%%%%%%%%%%%
%                         Simulations				    %
%%%%%%%%%%%%%%%%%%%%%%%%%%%%%%%%%%%%%%%%%%%%%%%%%%%%%%%%%%%%%%%%%%%%%%%%
\section{Simulation Details}\label{Computation}

The $N$-body simulations presented in this and the companion papers were
carried out using the open source adaptive mesh refinement code \mlapm\
(Knebe et~al. 2001). \mlapm\ reaches high force resolution by
refining high-density regions with an automated refinement algorithm.  
These adaptive meshes are recursive: refined regions can themselves 
be refined,
each subsequent refinement having cells that are half the size of the
cells in the previous level.  This creates a hierarchy of refinement
meshes of different resolutions covering regions of interest.  The
refinement is done cell-by-cell (individual cells can be refined or
de-refined) and meshes are not constrained to have a rectangular (or any
other) shape. The criterion for (de-)refining a cell is simply the number
of particles within that cell and a detailed study of the appropriate
choice for this number can be found elsewhere (Knebe et~al. 2001). The
code also uses multiple time steps on different refinement levels where
the time step for each level is a factor of two smaller than the time step
on the previous level. The latest version of \mlapm\ also includes an
adaptive time stepping that adjusts the actual time step after every major
step to restrict particle movement across a cell to a particular fraction
of the cell spacing, hence, improving the accuracy and computational time.

We first created a set of four independent initial conditions at redshift
$z=45$ in a standard \LCDM\ cosmology ($\Omega_0 = 0.3,\Omega_\lambda =
0.7, \Omega_b h^2 = 0.04, h = 0.7, \sigma_8 = 0.9$). Next, $512^{3}$ particles
were placed in a box of side length 64\hMpc\ giving a mass resolution of
$m_p = 1.6 \times 10^{8}$\hMsun.  For each of these initial conditions we
iteratively collapsed the closest eight particles to one particle reducing
our particle number to 128$^3$ particles. These lower mass resolution
initial conditions were then evolved until $z=0$.
  
At $z=0$, eight clusters from our simulation suite were selected in the
mass range  1--3$\times 10^{14}$\hMsun, each sampling differing environmental
conditions. Then, as described by Tormen \ea (1997), for each cluster the
particles within two times the virial radius were tracked back to their
Lagrangian positions at the initial redshift ($z=45$). Those particles 
were then
regenerated to their original mass resolution and positions, with the next
layer of surrounding large particles regenerated only to one level (i.e. 8
times the original mass resolution), and the remaining particles were left 64
times more massive than the particles resident with the host
cluster. This conservative
criterion was selected in order to minimise contamination of the final
high-resolution halos with massive particles.

\noindent

At the end of the high-resolution re-simulations the force resolution
is determined by the highest refinement level reached. The whole
computational volume was covered by a regular domain grid consisting of
256$^3$ cells. We had two separate criteria for refinement, a domain
cell was refined when there was more than one particle per cell,
further, every subsequent refinement was refined when there was more
than four particles per cell.  Thus the finest grid at $z=0$ consisted
of 65,536 cells per side, giving a force resolution of
$\approx$2\hkpc\ which allows us to resolve the host halos down to the
central $\sim$0.25\% of the virial radii of the host halos (see
\Table{HaloDetails}).

The halos chosen were selected to investigate the evolution of
satellite galaxies and their debris in an unbiased sample of host
halos, exploring the influence of environment upon the evolution of
such systems. To achieve this goal, excellent temporal resolution is
required - as such we retained 17 outputs from $z=2.5$ to $z=0.5$,
equally spaced with $\Delta t \approx 0.35$Gyrs, supplemented with an
additional 30 outputs spanning $z=0.5$ to $z=0$ with $\Delta t \approx
0.17$Gyrs. As we show in a companion paper (Gill~\ea 2004), the average number of orbits for our satellites is of the order 1-2. Therefore we have approximately 10-20 outputs available to define the orbit of a satellite, which is more that adequate to follow a live orbit properly. We found that to sufficiently sample a live satellite orbit you need at least eight time-steps. As you increase the time sampling the stability of the result quickly converges.

A simple analysis of the simulation at redshift $z=0$ provides us with
the relevant information on the host halo. At $z=0$ the halo masses
range from 1--3 $ \times 10^{14}$ \hMsun\ where the mass was defined
to be the total mass within the virial radius \Rvir, double counting both substructure and sub-substructure. The virial radii
in turn were defined at the point where the mean averaged density of
the host (measured in terms of the cosmological background density
$\rho_b$) drops below $\Delta_{\rm vir}=340$ with \Mvir\ being the
mass enclosed by that sphere. We then follow Lacey~\& Cole (1994) and
use their definition for formation time: the formation redshift
\zform\ is the redshift where the halo contains half of its present
day mass. Applying this criterion to our data we find that the ages of
our host halos have a spread ranging from roughly 8.3 Gyrs to as young
as 3.4 Gyrs. This alone shows that we are dealing with dynamically
different systems even though their masses are comparable; our older
halo's substructure has nearly twice the time to relax than the
youngest one's satellites. A summary of the eight host halos is
presented in Table~\ref{HaloDetails} where the halos are presented and
numbered from oldest to youngest. The variation in the number of
satellites from halo to halo with a trend for smaller hosts to contain
less satellites can be accounted for by the mass cut applied to the
satellites; as we expect a shift in the substructure mass function to
lower masses for smaller hosts we are artificially cutting off
satellites by applying a constant lower mass limit of 10$^10$\hMsun.

\noindent

%%%%%%%%%%%%%%%%%%%%%%%%%%%%%%%%%%%%%%%%%%%%%%%%%%%%%%%%%%%%%%%%%%%%%%%%
%                        MLAPM Halo finder				     %
%%%%%%%%%%%%%%%%%%%%%%%%%%%%%%%%%%%%%%%%%%%%%%%%%%%%%%%%%%%%%%%%%%%%%%%%

\section{\mhf: \mlapm's \texttt{H}alo \texttt{F}inder} \label{mhf}

The general goal of a halo finder is to identify gravitationally bound
objects. As all halos are centered about local over-density peaks they are
usually found simply by 
using the spatial information provided by the particle distribution.
Thus, the halos are located as peaks in the density field of the
simulation. To locate objects in this fashion, the halo finder is required
in some way to reproduce the work of the \nbody\ code in the calculation
of the density field or the location of its peaks. When locating halos
like this, the major limitation will always be the appropriate
reconstruction of the density field. With that in mind we introduce
\mlapm's-\texttt{H}alo-\texttt{F}inder, \mhf\ (or simply \finder)
hereafter.

\begin{figure}
   \centerline{\psfig{file=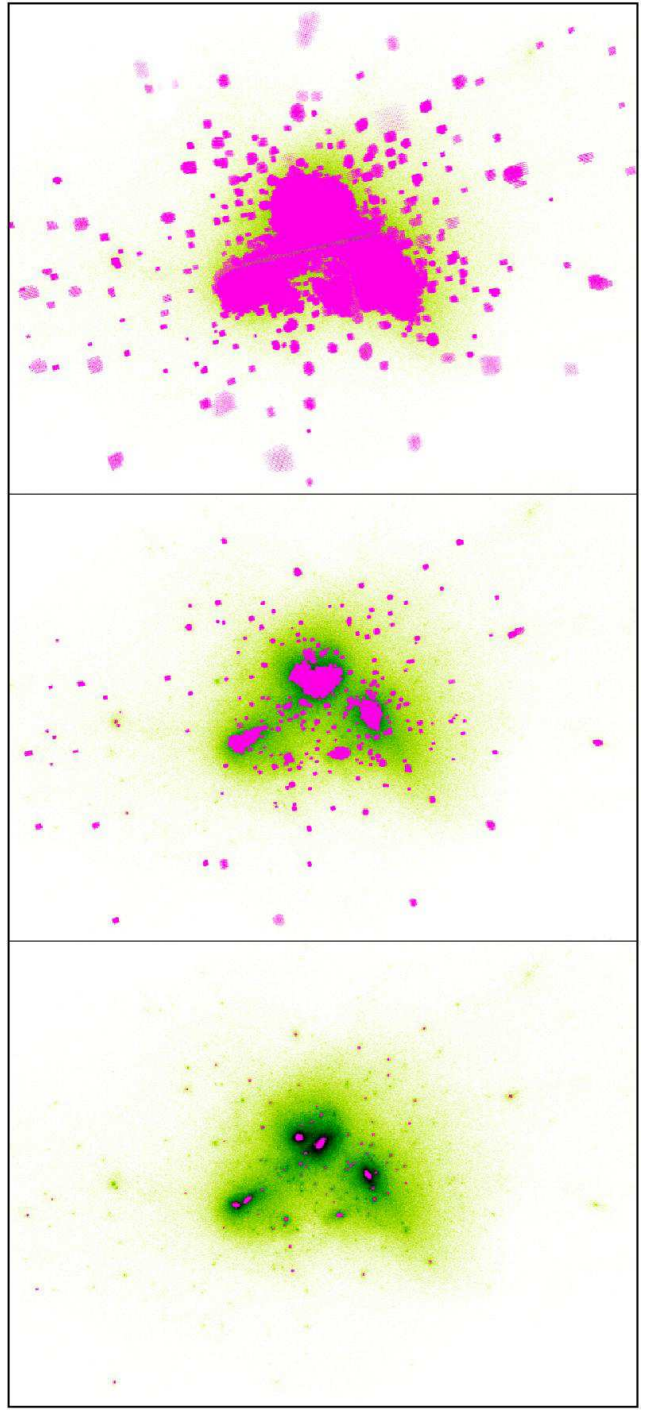,width=\hsize}}
   \caption{This panel shows a series of 3 consecutive refinement
   levels of \mlapm 's grid structure starting at the 5th refinement
   level superimposed upon the density projection of the particle
   distribution.}  
   \label{refgrid}
\end{figure}

\mhf\ essentially uses the adaptive grids of \mlapm\ to locate the
satellites of the host halo. As previously mentioned in
Section~\ref{Computation}, \mlapm's adaptive refinement meshes follow the
density distribution \it by construction\rm. Grid structure naturally
``surrounds'' the satellites, as the satellites are simply manifestations
of over-densities within (and exterior) to
the underlying host halo, a view which can best be appreciated through
inspection of \Fig{refgrid}. In this
figure, the refinement grids of \mlapm\ are superimposed over the
projected density of the particle distribution. The top image is the
$5^{\rm th}$ refinement level, with the $6^{\rm th}$ and $7^{\rm th}$ levels
shown below. We emphasise that the grids get successively smaller
and are subsets of other grids on lower refinement levels. The advantage
of reconstructing and using these grids to locate halos is that they
naturally follow the density field with the \textit{exact} accuracy of the
\nbody\ code. No scaling length is required, in contrast with techniques
such as FOF.  Therefore, \mhf\ avoids one of the major complications inherent
to most halo finding schemes as a natural consequence of its construction.

To locate appropriate halos within our simulation outputs we first
build a list of ``potential centres'' for the halos. Using the full adaptive
grid structure invoked by \mlapm, with the same refinement criterion as
for the original runs, we restructure the hierarchy of nested isolated
\mlapm\ grids into a ``grid tree'' and generate a list of prospective halo
centres by storing the centroid of the densest grid at the end of each
grid tree's ``branch''. 
Assuming that each of these peaks in \mlapm's adaptive grids is
the centre of a halo, we step out in (logarithmically spaced) radial bins
until the density reaches $\rho_{\rm satellite}(r_{\rm vir}) = \Delta_{\rm
vir}(z) \rho_b$(z), where $\rho_b$ is the universal background density,
unless we reach a point $r_{\rm trunc}$ where an upturn in the radial
density profile is detected.  This rise is encountered for (almost) all
satellites embedded within the background density of the host halo, a
point that we will discuss in more detail in Section~\ref{analmhf}.
The outer radius of the satellite is defined to be either $r_{\rm vir}$ or
$r_{\rm trunc}$, whichever is smaller, and dubbed \rmhf. Using all
particles interior to \rmhf\ we calculate other canonical properties for
each halo such as its mass, rotation curve, and velocity dispersion.

We now, however, need to prune the list of (still prospective) halos by
removing gravitationally unbound particles and duplicate halos.  The latter
occurs in two steps - first, for each
satellite a set of ``duplicate candidates'' is constructed based on the
criterion that their centres lie within each others' outer radii \rmhf.
Second, this list is then checked by comparing the internal properties
of the candidates. A candidate was affirmed to be a duplicate once its
mass, velocity dispersion, and center of mass velocity vector agreed to
within 80\%. We then kept the halo with the higher central density and
removed the other one from the satellite catalogue completely. 
This is a rare circumstance, yet one to which we will return in
\Sec{analmhf}. With our nearly complete set of halos now in hand, we
proceed to remove gravitationally unbound particles. This again is done in
an iterative process.  Starting with the \mhf\ halo centre, we calculate
the kinetic and potential energy for each individual particle in the
respective reference frame and all particles faster than two times the
escape velocity are removed from the halo. We then recalculate the centre,
and proceed through the process again.
This pruning is halted when a given halo holds fewer than eight
particles or when no
further particles need to be removed. We finish by recalculating the
internal properties of the halos with the radial density profiles of the
satellites fitted to the functional form proposed by Navarro,
Frenk~\& White (1997; hereafter, NFW)

\begin{equation}\label{NFWfit}
 \rho^{\rm cum}(r) =       \frac{M(<\!r)}{\frac{4\pi}{3}r^3}
                   \propto \frac{1}{(r/r_s) (1+r/r_s)^2} .
\end{equation}

\noindent
in the range from 8\hkpc\ ($\approx 4\times$force resolution) to
\rmhf. The scale radius $r_s$ is used to define the concentration of
the halo 

\begin{equation}\label{c}
 c=r_{\rm vir}/r_s.
\end{equation}

The procedure outlined above naturally deals with overlapping halos
and substructure halos, respectively.  But as mentioned before, for
such objects the virial radius can not be determined properly as we
will observe a rise in the radially binned density profile due to the
overlap with another halo or the embedding into the host. In that case
we set the outer radius of the (sub-)halo to be that point where the
density profile rises and all canonical properties are derived using
all (gravitationally bound!) particles interior to that radius. And
the fit to an NFW profile~\Eq{NFWfit} is only done out to that radius,
too. The situation is different once both of the overlapping halo's
centres are within each other's virial/upturn radius: we then checked, if
those two objects are just duplicates by comparing their internal
properties.

As stated in \Sec{Computation}, it is our aim to investigate the evolution
of satellite galaxies within their host halos. Thus, we restrict our
satellites to having at least 50 high-resolution simulation particles,
which corresponds to a mass-cut of $M_{\rm cut} \approx
10^{10}$\hMsun. Moreover, each satellite must contain at least 50\%
percent of its mass in high-resolution particles. In practice, this latter
constraint is not a critical one, relevant only for satellites beyond
twice the host halo's virial radius.

We can further take advantage of the \mlapm\ grids for measuring the
triaxiality of regions surrounded by an isodensity contour.
Essentially the various refinement levels are cuts in the density
field (isodensity surfaces). We calculated the inertia tensor for each
isolated refinement, weighting each cell by its density. Then using
the eigenvalues of the inertia tensor we construct the triaxiality
parameter (Franx, Illingworth~\& Zeeuw 1991) 

\begin{equation}
 T=(a^2-b^2)/(a^2-c^2) \ .
\end{equation}

To describe the host halo's triaxiality we used the $6^{\rm th}$
refinement level in
\mlapm. According to the refinement criterion adopted in the simulations
the $6^{\rm th}$ level surrounds material about 3000 times denser than
$\rho_b$ or, in other words, nine times denser than the material at the virial
radius.  A density of roughly $9\times \rho(r_{\rm vir})$ corresponds to
approximately the half-mass radius of the host.

\mhf\ is implemented into \mlapm\ in a way that provides the user
simultaneously with a snapshot of the dark matter particles \textit{and}
halo catalogues at each required output.  The most obvious advantages of
having the analysis performed ``on the fly'' are the reduction in computer
and human hours in the initial halo analysis stage. Embedding the halo
analysis in the code also enables us to potentially analyse the data at
unprecedented time resolution, if required.\footnote{\mhf\ can be switched
on either to act only when writing an output file (\texttt{-DMHF}) or at
each individual time step (\texttt{-DMHFstep}).} However, \mhf\ can also
be used with any already existing single time-step snapshot and hence is
not limited to data produced by \mlapm; it can also be used for any \nbody\
output provided the latter is converted to \mlapm's binary format using the
tools included in the \mlapm\ distribution.

%%%%%%%%%%%%%%%%%%%%%%%%%%%%%%%%%%%%%%%%%%%%%%%%%%%%%%%%%%%%%%%%%%%%%%%%
%                        Halo finder Analysis				%
%%%%%%%%%%%%%%%%%%%%%%%%%%%%%%%%%%%%%%%%%%%%%%%%%%%%%%%%%%%%%%%%%%%%%%%%
\section{Analysis of \mhf\ halos}\label{analmhf}

\mhf\ was applied to each of the 376 temporal outputs (47 outputs per 
each of the eight independent halos), providing us with a list of
all satellites and their internal properties at each individual redshift
under consideration. As stated earlier, the detailed analysis of the
science associated with this study is presented in Paper~II (\GKGD).  
We do however wish to highlight several key preliminary results here
which relate specifically to the halo identification process.

\begin{figure}
   \centerline{\psfig{file=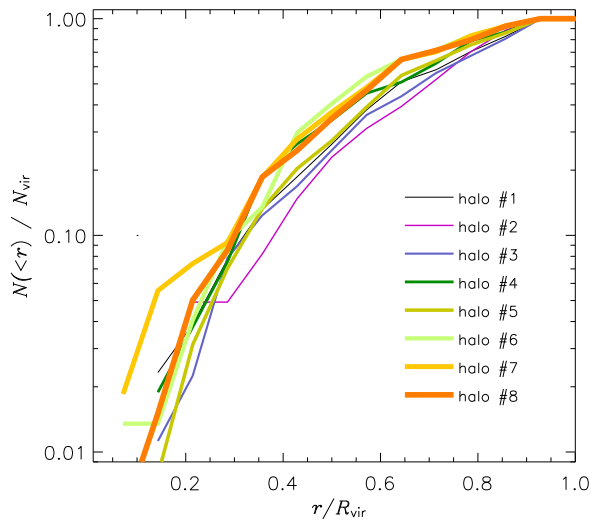,width=\hsize}}
   \caption{Number of satellites (normalised) 
	    orbiting within the virial radius of the 
            host halo at $z=0$, as a function of radial distance $r$, 
            normalised by the virial radius of the respective host \Rvir.}  
   \label{Nr1}
\end{figure}

In \Fig{Nr1} we plot the normalised number of satellites as a
function of normalised radius. One, perhaps not surprising, aspect of
Figure~\ref{Nr1} is the similarity in the slopes. Although the number of
satellites in each halo may vary, the relative radial distribution of
the satellites is similar across halos. This is reminiscent of the
universal density profile of dark matter halos, as described by
NFW.  Although the radial distribution of the satellites
remains consistent, there exists a range of substructure densities for the
halos, as there is a spread in the number of satellites within each
halo (recall Table~1). 
Therefore, we should be able to distinguish the effects of
substructure density on the physical properties of the satellites.

The other striking feature of Figure~2 is the lack of satellites in the
inner 15\% of the virial radius.  One might ask why this is the case.
Does it indicate
that satellites dynamically avoid the central region of the
halo?  Perhaps they simply do not spend much time there?  Perhaps via
physical means satellites that venture near the centre are either merged
or experience such strong tidal forces that they are destroyed?  
Perhaps we are simply dominated by numerical effects and are 
witnessing the premature
destruction of halos in dense environments. This latter problem - known as
\textit{overmerging} - affected low-resolution dissipationless simulations 
in the early 1990s,
failing to produce galaxy-sized dark matter halos in clusters
(e.g. Summers \ea\ 1995; van~Kampen 1995; Moore \ea\ 1996).  Traditionally,
this was explained by the lack of dissipation in the simulations.
With the inclusion of a baryonic
component, denser objects could form and survive in the centres of
these dense regions. However, with the onset of higher resolution
simulations a converse effect was encountered - specifically,
an \textit{abundance} of substructure was found 
(Klypin~\ea 1999).

The explanation of overmerging (or substructure disruption) was accredited
to numerical limitations in the simulations. van~Kampen (1995) found that
particle evaporation due to two-body effects is only important for low
particle number halos ($<$30 particles ). Moore~\ea (1996) further
investigated particle halo heating, which they concluded was 
negligible should 
sufficient mass resolution exist. Moore~\ea also demonstrated the for
satellites in a static host potential, if a simulation had insufficient
spatial resolution, halos would have artificially large cores and hence 
undergo accelerated tidal disruption.
They also found that halos become unstable and
are erased when the tidal radius is smaller than approx 2-3 times the halo
core radius (which itself can be related to the gravitational softening 
length).
Klypin \ea (1999) investigated the issue of ``overmerging'' in great detail
using a variety of higher resolution simulations, concluding that the
resolution required to avoid artificial destruction of galaxy-sized halos
of mass $\approx 10^{11}$\hMsun was $\leq 2$ \hkpc\ (spatial) and $\leq
10^9$\hMsun\ (mass).

Since we appear to have sufficient numerical resolution and our
data lies well within the limits of not being dominated by overmerging,
one might query whether or not 
the lack of substructure in the inner region is due to a
limitation of our halo finder.  When defining the radius of our halos we
could not for all halos follow the density profiles out to \rvir\ defined
via $\rho_{\rm sat}(r_{\rm vir}) = \Delta_{\rm vir} \rho_b$; as noted
earlier, it was
necessary in many cases 
to define a truncation radius $r_{\rm trunc}$. The existence of
$r_{\rm trunc}$ generally indicates that the satellite is embedded within
the host's density field, as already noted by Bullock \ea (2001).
Thus, as a satellite gets closer to the central density region of the host
halo, its overdensity peak becomes less contrasted. It is intrinsically
harder to find satellites with low central densities under the standard
paradigm of halo finding, especially close to the cuspy centre. It is not
at all obvious how to disentangle the particle distribution of the
satellite and the host halo: this is a fundamental limit to finding halos
in the traditional way of observing over-densities and requires further
investigation. In the next section, however, we introduce a method of
finding halos that eliminates the background halo and, hence, minimises
this problem.

\begin{figure}
   \centerline{\psfig{file=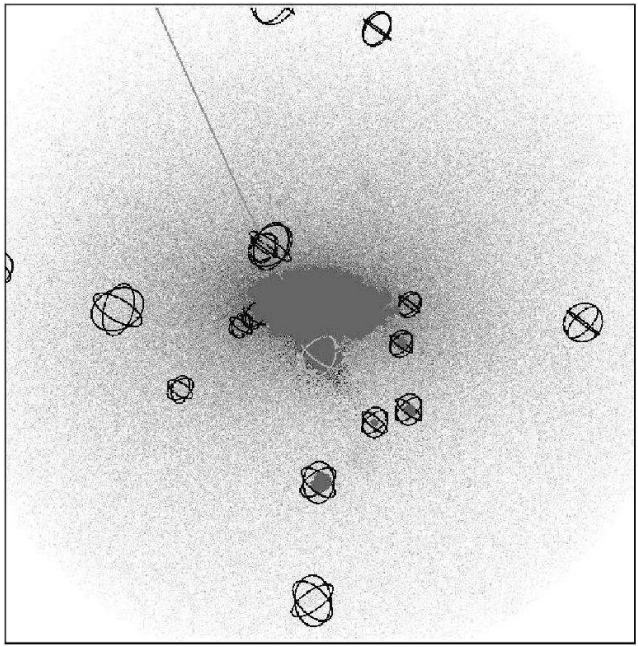,width=\hsize}}
   \caption{The inner 250 \hkpc\ of halo \#1 with the particles' line of 
            sight density shown.  We show the grids of the 7$^{th}$ refinement 
            level of \mlapm, with the central refinement about 60\hkpc\ in 
            radius. The dark spheres represent the satellites located by \mhf. 
            The light sphere surrounds a satellite galaxy not found by
	    \mhf. The apparent sizes of the spheres are simply a visualisation
            effect as spheres farther away from the virtual observer appear
            smaller.}
   \label{mhf-badies}
\end{figure}

The \mhf\ method fails in the inner regions for two reasons. Firstly,
because it is hard to detect the upturn in the density field, 
substructure is eliminated through suspected duplication of a halo because
the substructure's radius has been falsely tracked out to essentially the
virial radius of the host, its own upturn radius has been missed. The second reason results from a fundamental
flaw in \mhf's methodology - that the smaller satellite grids
merge with the host's refinement grid and hence do not produce an isolated
refinement. Therefore we are losing potential centres, a problem 
illustrated further in \Fig{mhf-badies}.  There, we show the inner
250\hkpc\ of halo \#1, along with the grids for the 7$^{th}$ refinement level
of \mlapm\ (gray-shaded areas) with the central refinement about 60\hkpc\ in
radius. The dark spheres indicate the positions of satellites located by
\mhf. Note that those dark spheres that do not encompass an isolated
refinement grid would do so at one of the next coarser (or finer) levels.
The light sphere at the border of the ellipsoidal host refinement also
surrounds a satellite galaxy. However, this object was \textit{not} picked
up by the \finder\ but rather by the \tracker\ outlined in the next
section. \mhf\ was unable to identify this satellite as an individual
object as its refinement grid has merged with the host's grid, thus not
allowing an isolated refinement and a potential center, respectively. The
straight line pointing to this satellite is simply its orbital path.

\begin{figure}
\centerline{\psfig{file=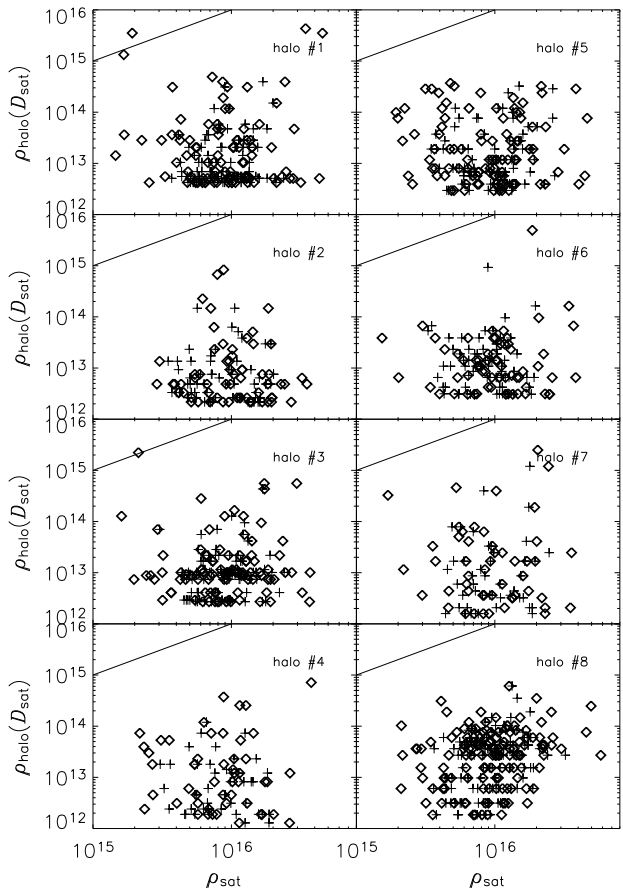,width=\hsize}}
   \caption{The density of the host halos at the radial distance \Dsat\ of the 
            satellite versus the central density of the satellites. The crosses
            represent the satellites found by \finder\ while the diamonds are 
            the satellites found by \tracker.}  
   \label{DensDens}
\end{figure}

The problem can be viewed differently in \Fig{DensDens}. Here we plot the
radially averaged density of the host halo at the position of a satellite
against the maximum, central density of the satellite itself for redshift
$z=0$. The results are presented for the \finder\ (crosses) as well as for
the \tracker\ (diamonds) to be introduced in \Sec{mht}. The line running
through the plot for each individual host halo marks the 1:1
correspondence: satellites that fall onto (or even above) this line have
central densities equal to (or smaller than) the host environment they are
embedded within. We do observe a general (and reasonable) trend for satellites
to have higher central densities than their dark matter vicinity. However,
the figure also proves that the \tracker\ tends to also find satellites
less contrasted and closer to the 1:1 relation, respectively.  When
interpreting \Fig{DensDens}, and especially comparing the \finder\ to the
\tracker\ results, one needs to bear two things in mind: firstly, there
are many more \tracker-satellites obscuring a one-to-one comparison with
\finder, and secondly, \finder\ relies on \rmhf\ as the final point of the
profile whereas \tracker\ has the ability to properly measure \rvir. Minor
differences in binning can also 
lead to very small changes in the central density calculation.

%%%%%%%%%%%%%%%%%%%%%%%%%%%%%%%%%%%%%%%%%%%%%%%%%%%%%%%%%%%%%%%%%%%%%%%%
%                  MLAPM Halo tracer					     %
%%%%%%%%%%%%%%%%%%%%%%%%%%%%%%%%%%%%%%%%%%%%%%%%%%%%%%%%%%%%%%%%%%%%%%%%
\section{\mht: \mlapm's \texttt{H}alo \texttt{T}racker}\label{mht}

Conventional halo finders have a rich history in identifying isolated
systems.  In this regard, \mhf\ might be viewed as simply an
alternative approach to an already reasonably well-understood
problem. To be fair though, \mhf\ does push the conventional paradigm
of simply using the three dimensional spatial data to locate the halos
to the limit, locating the halos with (nearly) the exact accuracy of
the \nbody\ code. Having this ability, \mhf\ becomes the ideal halo
finder to locate substructure for \mlapm\ and no doubt an excellent
halo finder for other codes. Although, as we have seen in
Section~\ref{analmhf}, apart from numerical limitations in \nbody\
codes (e.g. overmerging) there still remain limitations to our current
halo finders. These limitations become a problem in the simulations
when considering the substructure of any dense system, for example
galaxy clusters and galaxies. Therefore, to successfully find the
substructure we need to change the paradigm used to find it.

To successfully make this change we must first understand the environment
in which we are finding the substructure and then exploit its
characteristics. One characteristic is that most halos conserve their
identities, that is substructure halos rarely undergo mergers in halos
because of their high relative velocities (Ghigna \ea 1998; Okamoto \&
Habe 1999). Further, halos in dense environments undergo tidal stripping
and substructure interactions, no longer accreting material, but being
stripped of it. Thus in such environments it is sufficient to trace the
particles of the satellite once the satellite has entered the host's
virial radius. In this section we introduce
\mlapm's-\texttt{H}alo-\texttt{T}racer, \mht\ (or simply \tracker)
hereafter.

\mht\ takes an arbitrary output of our new \mlapm-based halo finder \mhf,
and correlates the particles for an arbitrary number of time steps
using all simulation outputs from that initial output until redshift
$z=0$. In our particular investigation it was appropriate to define
that initial arbitrary output to be the formation time \zform\ of the
host halo or the time when the host halo contained half of its present
mass (Lacey \& Cole 1994). We then followed all the satellites that
were within two times the virial radius of the host halo at this
formation time. Although, we miss a few satellites due to \mhf's
identification limitations in the inner 10-15\% of the halo, from this
time on \mht\ precisely follows the orbits of our initial set of
satellites irrespective how close they come to the host's
centre. Explicitly, \mht\ takes the particles from an initial \mhf\
analysis and then locates these particles in the next available output
again. \tracker's first task is then to (re-)calculate the halo's
centre.  This is done by using the centre-of-mass of the innermost 20
particles from the previous time step as an initial estimate, then
using the same iterative method to check the credibility of the halo,
as outlined in Section~\ref{mhf}. Once the satellite was identified as
bona fide the radial profile was generated, a NFW profile fitted, and
other canonical properties calculated. The binning for the profile
again used logarithmically spaced radial bins covering the entire
particle distribution.  The radius of the satellite was consistently
determined as being the radius when the cumulative density profile
dropped below $\rho_{\rm satellite}(r_{\rm vir}) = \Delta_{\rm vir}(z)
\rho_b(z)$. This time we will not encounter the situation where the
profile rises again as is the case for \finder; the satellite is no
longer embedded within the ``particle background'' of the host halo
but treated as a separated entity.

There are a number of advantages in tracing the halos in this way -
first, because we are tracking just the satellites' individual
particles we do not have the complication of the background density
distribution and the consequent lack of contrast against the host
system for all outputs $z<z_{\rm form}$. Following from this, we do
not have to accept the truncation radius - the radius where the \finder\ encounters
and upturn in the denisty profile - as the ``natural'' radius of
the satellit. Further, this
method allows us to investigate the development of tidal streams,
which forms the basis of the extensive analysis provided in Paper~III.

%%%%%%%%%%%%%%%%%%%%%%%%%%%%%%%%%%%%%%%%%%%%%%%%%%%%%%%%%%%%%%%%%%%%%%%%
%                        Halo Tracker Analysis				%
%%%%%%%%%%%%%%%%%%%%%%%%%%%%%%%%%%%%%%%%%%%%%%%%%%%%%%%%%%%%%%%%%%%%%%%%

\section{Analysis of \mht\ halos}\label{analmht}

The nature of hierarchical structure formation, i.e. mergers, dynamical
and tidal destruction of substructure, requires a little more work when
applying \mht\ to our simulation data. Even though were are now tracing
the initially bound particles forward in time, we need a criterion to
decide whether a satellite galaxy is disrupted or still alive. We
therefore introduce the tidal radius as given below.

\subsection{Tidal Radius}\label{TidalRadius}

The tidal radius is defined to be the radius of the satellite where the
gravitational effects of the host halo are greater than the self-gravity
of the satellite. When approximating the host halo and the satellite as
point masses and maintaining that the mean density within the satellite
has to be three times the mean density of the host halo within the
satellites distance $D$ to the host halo (Jacobi limit) the definition for
tidal radius reads as follows

\begin{equation}\label{TidalRad}
 r_{\rm tidal} = \left( \frac{m}{3M} \right)^{\frac{1}{3}} D \ ,
\end{equation}

\noindent
where $m$ is the mass of the satellite and $M=M(<D)$ is the mass of
the host halo internal to the distance $D$.

In order to stabilise the determination of the tidal radius \rtidal\ we
actually use an iterative procedure again. By defining the satellite mass
$m$ as the mass internal to \rtidal, i.e. $m=m(<r_{\rm tidal})$, we find
\rtidal\ by solving

\begin{equation}\label{TidalRadEQ}
 r_{\rm tidal} - \left(\frac{m(<r_{\rm tidal})}{3M}\right)^{3} D = 0 \ .
\end{equation}

\noindent
Starting with the distance of the furthest particle in the satellite
particle distribution as the initial guess for \rtidal, the method quickly
converges in only two-to-three iterations.

\subsection{Satellite Disruption}

As the satellites orbit within the host halo they undergo tidal stripping,
hence, the satellites are gradually losing mass. Therefore, their
particle distribution becomes more and more diffuse, reducing the tidal
radius of the satellite. Eventually there comes a point when the satellite
loses sufficient mass that we begin to reach the limits of our numerical
resolution, not having sufficient number of particles to follow the
satellite further. The satellite might have survived for longer, however,
we do not have the resolution to follow it. Thus when a satellite passes
through our numerical limit we tag it ``disrupted''. In practice this means
that if there are fewer than 15 particles within the tidal radius we
classify the satellite as being disrupted. Note that we are unable to
separate numerical resolution disruption and real physical disruption of a
satellite. It is not clear if we had infinite mass resolution that the
satellite would still actually survive.

We also stress that since our tidal radius formulae assumes circular
orbits, the usual $D$ corresponds to the pericentric distance (Hayashi
\ea 2003); however, we calculate \rtidal\ for each satellite at each
individual output to check for (tidal) disruption.

Furthermore, particles outside the tidal radius are not automatically
stripped from the halo - what is just as important is the time spent
under the influence of that tidal field, it still takes time for the
particle to climb out of the potential. For example a satellite might
be on a very eccentric orbit and pass close to the centre of the host
halo, thus having a very small tidal radius at the pericentre. Now it
is true that most of the tidal stripping will occur at this
pericentre, however, because the satellite only spends a short time
there not all the particles outside the tidal radius will be stripped.

\subsection{Orbital Information}

\begin{figure}
   \centerline{\psfig{file=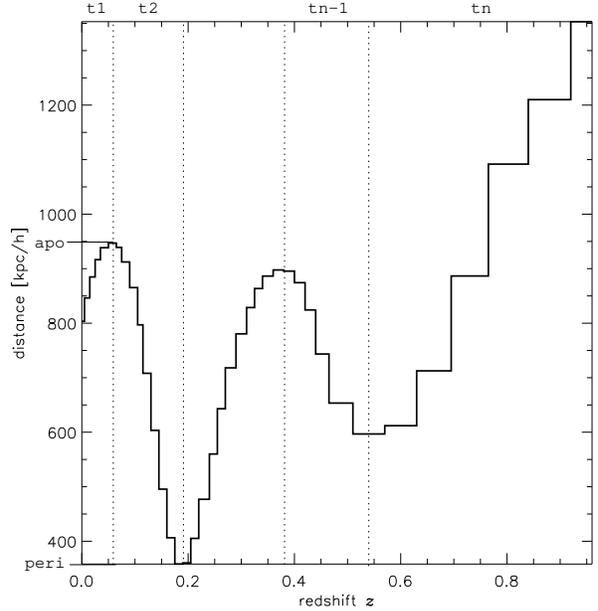,width=\hsize}}
   \caption{Distance from the center of the host halo in real co-ordinate
            system  as a function of redshift for one particular satellite
            with ($M_{\rm sat}/M_{\rm host}=0.7\times 10^{-2}$).}  
   \label{orbit}
\end{figure}

In \GKGD\ we present a detailed analysis of the satellites' orbits, but 
do take the opportunity here to provide some basic terminology relevant to
both Papers~I and II here.
The high temporal resolution of our simulations ($\Delta
t=0.17$ and $\Delta t=0.35$ Gyrs, respectively, for $z>0.5$ and $z<0.5$) 
enables us to track in
detail the orbits of the satellites. As an example, in
\Fig{orbit} we show the orbit of one particular satellite. We can see that
this satellite initially plunged in from outside the virial radius at
$z=0.8$ and was subsequently captured by the host, undergoing two further
orbits prior to $z=0$. This orbital information was then used to 
construct a measure of eccentricity

\begin{equation}\label{ElipFunk}
 \epsilon = 1 - \frac{p}{a}
\end{equation}

\noindent
where $p$ is the most recent ``closest'' distance to the host's centre as a
minima (labelled \texttt{peri} in \Fig{orbit}) and $a$ the most recent
``furthest'' distance (labelled \texttt{apo}) as a maxima. Moreover, we are
also in the position to calculate the number of orbits from the time
evolution of the distance to the host centre. To this extent, we simply
count the number of extrema (four in the case shown) and divide by
two. We further correct for incomplete orbits at the beginning and end
points of the distance relation, resulting in the following formula for
the total number orbits:

\begin{equation} \label{norbits}
 N_{\rm orbits} = \frac{N_{\rm extrema}}{2} +
                  \min(\frac{1}{2},\frac{t_1}{t_2}) + 
                  \min(\frac{1}{2},\frac{t_n}{t_{n-1}})
\end{equation}

\noindent
The number of orbits measured by that method for the sample satellite
presented in \Fig{orbit} is $N_{\rm orbits}=2.69$.

We emphasise though that the orbits of the satellites are not always
as aesthetically ``smooth'' as that for the
one presented in \Fig{orbit}. We are dealing here with
live potentials and hundreds of satellites orbiting within it 
simultaneously.  The host halo is constantly growing in mass and shows internal
oscillations in shape due to ongoing mergers (see \GKGD). This has 
has an impact on the orbital evolution of the satellites, as described in
Paper~II.

\subsection{The radial distribution of satellites}
%-------------------------------------------------

\begin{figure}
   \centerline{\psfig{file=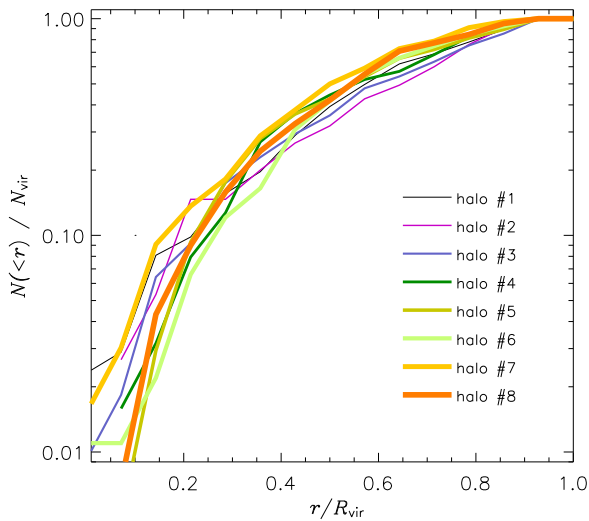,width=\hsize}}
   \caption{Number of satellites within a particular radius normalised by the
            total number of satellites as a function of radius normalised by
            the virial radius at $z=0$. Only satellites more massive than
            $2\times 10^{10}$\hMsun\ were taken into account.}
   \label{Nr2}
\end{figure}

In \Fig{Nr1} we showed the radial distribution of satellites for
\mhf. We now present in \Fig{Nr2} the same plot for \mht\ highlighting
the superiority of the
\tracker. Once again we observe the similarity in the slopes across the
eight halos, a so-called 
``universal satellite distribution''. However, the important
result is that using the \tracker\ we now find (more) objects within the
central 10\% of \rvir\ of the host halos. For halo~\#1, for instance, we
located 5 satellites with mass greater that $10^{10}$\hMsun\ within 10\%
of the virial radius with the closest satellite at $z=0$ being a mere
35\hkpc\ away from the host's centre.

\begin{figure}
   \centerline{\psfig{file=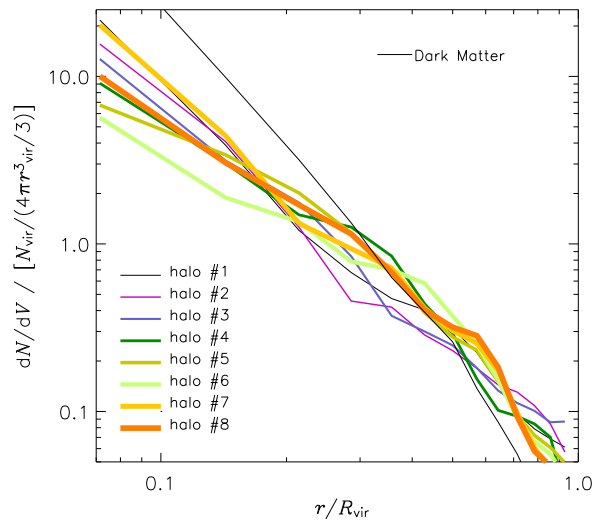,width=\hsize}}
   \caption{Number density of satellites normalised by the average number of
            satellites per unit volume as a function of radius normalised by 
            the virial radius at $z=0$. Only satellites more massive than 
            $2\times 10^{10}$\hMsun were taken into account.}
	\label{Nr3}
\end{figure}

To allow a more natural comparison to work published by other
authors, we present the data in a slightly different fashion in
\Fig{Nr3}. This time the radial number density of satellite
galaxies is shown. As with Ghigna~\ea (2000) and De Lucia~\ea (2003) we
also find that the subhalo population is ``anti-biased'' relative to
the dark matter distribution in the inner regions of the
halos. Moreover, we again observe no trend with environment for the
sample of eight halos under investigation; all halos, irrespective of
age and richness, do show the same anti-bias in the satellite
distribution.

%%%%%%%%%%%%%%%%%%%%%%%%%%%%%%%%%%%%%%%%%%%%%%%%%%%%%%%%%%%%%%%%%%%%%%%%
%                  COMPARISON
%%%%%%%%%%%%%%%%%%%%%%%%%%%%%%%%%%%%%%%%%%%%%%%%%%%%%%%%%%%%%%%%%%%%%%%%

%%%%%%%%%%%%%%%%%%%%%%%%%%%%%%%%%%%%%%%%%%%%%%%%%%%%%%%%%%%%%%%%%%%%%%%%
\section{Comparison to other halo finders}\label{compare}

\begin{figure}
   \centerline{\psfig{file=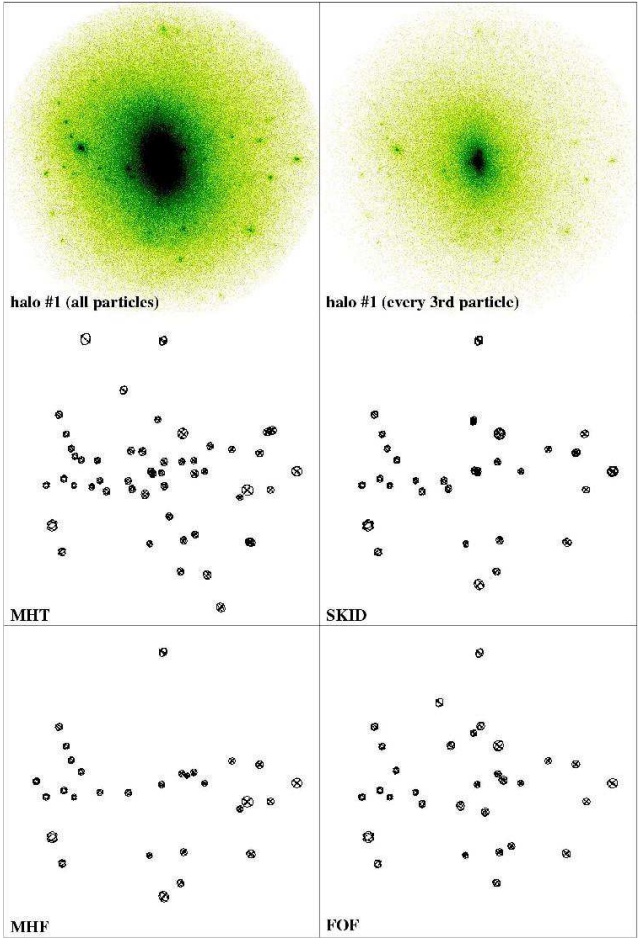,width=\hsize}}
   \caption{The top two panels show the line of sight density
   projection of particles for halo~\#1. The left panel displays all
   the particles, while the right panel shows every third particle.
   The remaining four panels show the results of the halo finding
   algorithms: \mht, SKID, FOF, and \mhf\ (clockwise from top left).
   A sphere of fixed radius 20\hkpc\ surrounds each satellite.}
   \label{FinderCompare}
\end{figure}

In this section we compare \mhf\ and \mht\ to two other halo finders,
namely SKID and FOF. In \Fig{FinderCompare} we present a visual comparison
of the effectiveness of the respective halo finders. Firstly, the top two
panels show the line of sight density projection of particles for halo~\#1
within a 700\hkpc\ sphere. The left panel displays all the particles,
while the right panel only shows every third particle. In the remaining
four panels, we present the results of the halo finding algorithms:
(reading clockwise, starting upper left) \mht, SKID, FOF, and \mhf. A
sphere of fixed radius 20\hkpc\ surrounds each located satellite where
different apparent sizes are simply visualisation effects, i.e. spheres
farther away from the virtual observer appear smaller.

\begin{figure}
   \centerline{\psfig{file=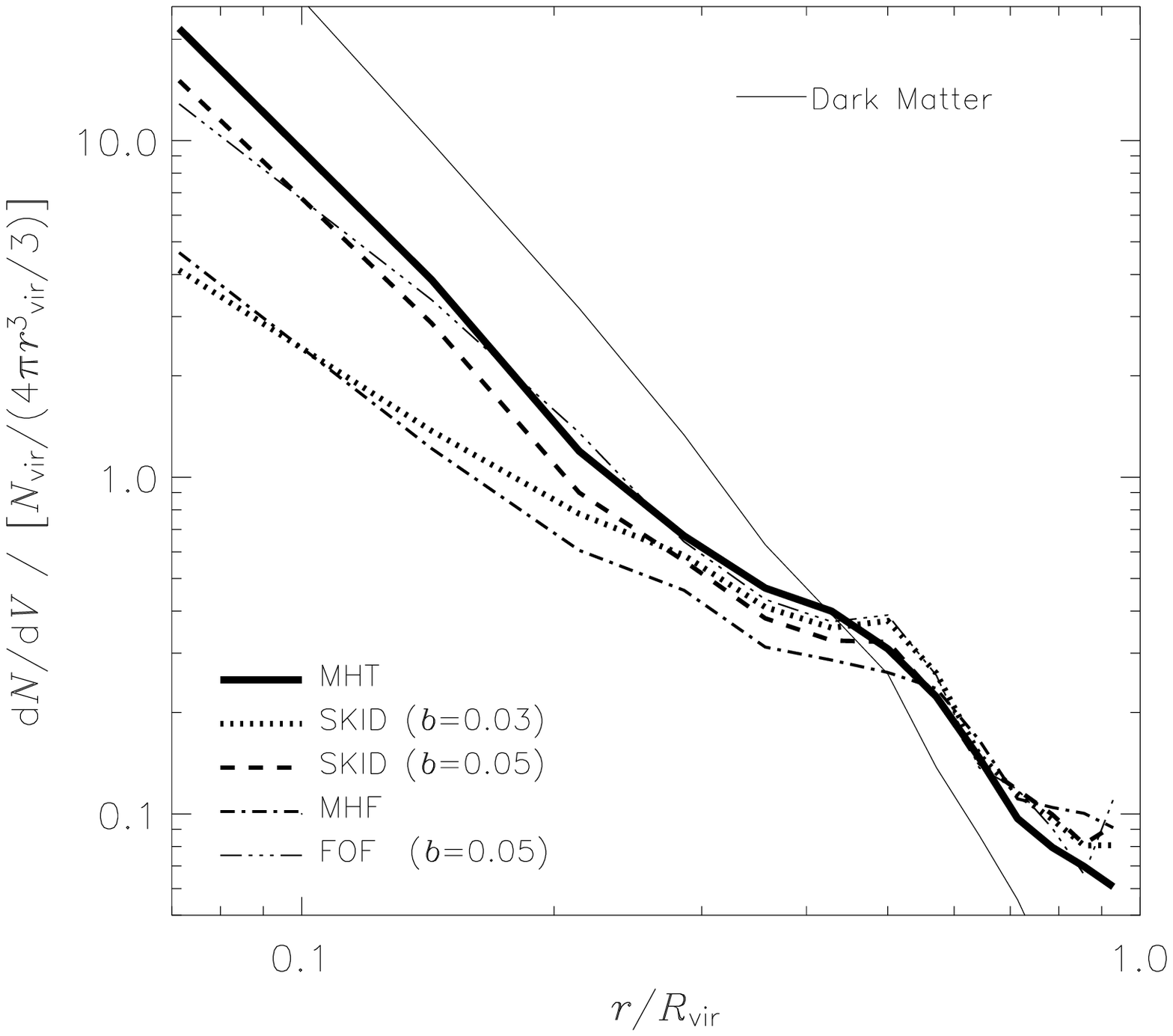,width=\hsize}}
   \caption{Number density of satellites normalised by the average number of
            satellites per unit volume as a function of radius normalised by
            the virial radius at $z=0$, halo \#1 for each of the halo finding
            algorithms: \mht, SKID, \mhf, FOF. Only satellites more massive 
            than $2\times 10^{10}$\hMsun were taken into account.}
   \label{NrComparedens}
\end{figure}

\begin{figure}
\centerline{\psfig{file=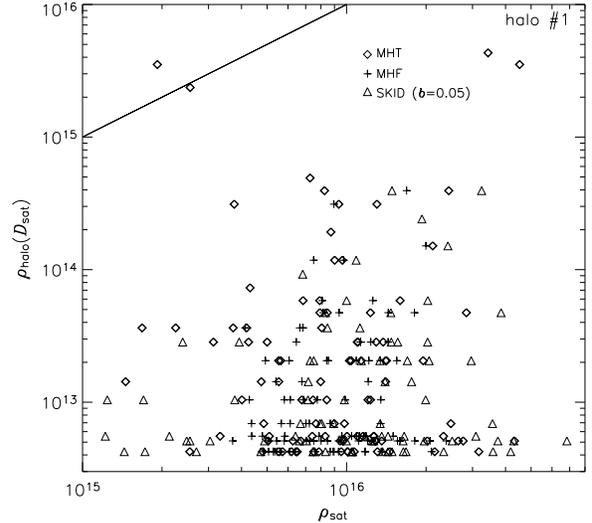,width=\hsize}}
   \caption{Same as \Fig{DensDens}, but this time comparing \tracker\
            (diamonds) and \finder\ (crosses) with the SKID, $b$=0.05
            (triangles) analysis for halo~\#1.}  \label{DensDensSKID}
\end{figure}

SKID was run with multiple linking lengths and the option to remove
gravitationally unbound particles enforced.\footnote{For a detailed
description of SKID please refer to
\texttt{http://www-hpcc.astro.washington.edu/tools/skid.html}.} Under
close visual inspection, the best SKID results were found with linking
lengths $b$=0.03 \& $b$=0.05 times the interparticle separation.  The
same values were taken for the FOF analysis. The results from the
analysis with linking length $b$=0.05\footnote{A linking length of
$b$=0.05 expressed in terms of the interparticle separation translates
into a physical linking length of $\sim$6\hkpc, which is roughly three
times the force resolution.} appear to be the most reliable ones and
hence are shown in \Fig{FinderCompare}. We need to stress though that
the analysis of the SKID and FOF results can be further refined by
combining various halo catalogues into a tree as explained in, for
instance, Ghigna~\ea (2000). However, one of the benefits our
\mlapm\ halo finders is that neither requires any further input such
as a linking length.  Moreover, the latest version of \mlapm\ is
capable of performing the analysis ``on-the-fly''.  A visual
inspection of \Fig{FinderCompare} indicates that \mht\ provides the
most complete halo list. Specifically, the \tracker\ found 53
satellites, the \finder\ found 32, SKID 33 and FOF 32 within the
plotted spherical region of diameter 1.4\hMpc. Within the sets of
satellites, there is of course considerable overlap. Essentially, each
set of satellites found by \mhf\, SKID, and FOF were subsets of
\mht. To look at this quantitatively, we calculate the radial number
density of satellites for all four halo finders again (cf.
\Fig{Nr3}). This time we concentrate on halo~\#1 highlighting the
differences between the halo finding methods. The result is presented in
\Fig{NrComparedens} which clearly shows the success of the \tracker\ over
all other methods. However, it is interesting to note that a simple FOF
analysis gives quantitatively similar results to the more sophisticated
SKID data. The difference between the \finder\ and the \tracker\
is quite remarkable, as is the similarity between \mhf\ and SKID with
$b$=0.03. Our explanation for the lack of substructure in the central
region for this particular SKID analysis is that these objects where
either removed because of our lower mass cut of 100 particles or the fact
that SKID did not classify them as gravitationally bound.  What we can also
learn from \Fig{NrComparedens} is that the sensitivity of the halo finder
in the inner regions can severely bias the results. For example using
\mhf\ or SKID with b=0.03 provides a much stronger anti-bias in the
satellite distribution than the more appropriate \tracker\ and SKID
($b$=0.05) analysis, respectively.

We like to the close the comparison by coming re-examining
\Fig{DensDens}, this time including the results from the SKID ($b$=0.05)
analysis. The result for halo~\#1 can be viewed in \Fig{DensDensSKID}.
We still observe that \textit{only} the \tracker\ is capable of
resolving satellites with central densities close to the (local)
density of the host halo. We also inspected the situation for the FOF
($b$=0.05) data and could not find any significant difference and
hence decided to not plot the data for clarity.

%%%%%%%%%%%%%%%%%%%%%%%%%%%%%%%%%%%%%%%%%%%%%%%%%%%%%%%%%%%%%%%%%%%%%%%%
%                          CONCLUSIONS                                 %
%%%%%%%%%%%%%%%%%%%%%%%%%%%%%%%%%%%%%%%%%%%%%%%%%%%%%%%%%%%%%%%%%%%%%%%%
\section{Conclusions} \label{Summary and Conclusions} \label{conclusions}

Computational cosmology is not only limited by crucial factors such as
the dynamical range and the mass resolution, but also by its
analysis tools. We emphasise, perhaps obviously to most readers, that
\nbody\ codes simulating
structure formation in the Universe can only ever be as useful as their
associated analysis tools allow.

In this paper we presented two new methods for identifying
gravitationally bound objects in such simulations. Both methods are
based upon the open source adaptive mesh refinement code \mlapm\
(Knebe et~al.  2001). They both exploit the refinement hierarchy of
said code and hence locate halos as well as halos-within-halos with
exactly the same accuracy as \mlapm\ simulates their evolution. We
showed the limitations of a simple snapshot analysis and how it can be
overcome by taking into account the whole history of each halo. Thus
not restricting the halo finding algorithm to just the spatial
information but rather including the velocity information as well,
hence, using the full six-dimensional information available.

Not only do we intend to implement the halo finder into the distribution of
\mlapm\ allowing for an on-the-fly analysis saving both computational
and human resources in the analysis process, but it is also our intention to 
have it as a stand alone program. In both cases, the implementation will 
be such that it can analyse a single output of any given \nbody\ code.

We showed that halo finding still possesses the inherent
problem of overmerging in the
very central regions of the host. However, by tracking satellites rather
than identifying them at separate time snapshots of the simulation we
learned that this problem is \textit{not} overmerging in the conventional
sense. The objects are in fact present and simulated properly, but their
densities have insufficient contrast to be picked up by a simple ``finder''.
Only when tracking them in time from (at best their very own) formation
time were we able to quantify their existence as close as 5\% of $r_{\rm
vir}^{\rm host}$ to the cuspy centre. We do not intend to question the
credibility of other (most excellent) halo finders such as SKID though. We
rather pointed out that the results in the central region are subject to
subtleties that can be most easily avoided by tracking satellites.

It has recently been pointed out by Taylor~\ea (2003) that the radial
distribution of the Milky Way's satellite galaxies does not reconcile
with the predictions of semi-analytical models of galaxy formation and
cosmological simulations, respectively, which has been confirmed by
Kravtsov~\ea (2004). Even though there is quite a prominent
substructure population in the simulations it is mostly clustered in
the outer regions of the host whereas the Milky Way satellites are
preferentially found closer in.  It remains unclear if this poses a
new challenge to the CDM structure formation scenario or simply
reflects what we presented in this study. Namely, identifying
substructure halos that lie close to the centre of the host is
intrinsically challenging and might be overcome by actually
\textit{tracking} the satellites rather than \textit{finding} them.

Finally, we further increase the statistics to suggest that anti-bias
of the satellite distribution is a common property of the satellite
distribution, perhaps even universal. However, this antibias is 
still most likely a relic of
numerical overmerging. A convergence study for the satellite
populations is critical in resolving this issue and reconciling the
lensing observations. However, one very interesting result is the
continued self-similarity in the dark matter distribution, the
satellite radial distribution is common, once again perhaps "universal"
 even though the number of
substructure satellites changes, reminiscent of the universal density
profile of the underlying host.

%%%%%%%%%%%%%%%%%%%%%%%%%%%%%%%%%%%%%%%%%%%%%%%%%%%%%%%%%%%%%%%%%%%%%%%%
%                         ACKNOWLEDGEMENTS                             %
%%%%%%%%%%%%%%%%%%%%%%%%%%%%%%%%%%%%%%%%%%%%%%%%%%%%%%%%%%%%%%%%%%%%%%%%
\section{Acknowledgments}
The simulations presented in this paper were carried out on the Beowulf
cluster at the Centre for Astrophysics~\& Supercomputing, Swinburne
University. We acknowledge the financial support of the Australian
Research Council, the Swinburne Research Development Grant scheme, and
Mike Dopita. Finally, we wish to thank Chris Power for helpful correspondences.

%%%%%%%%%%%%%%%%%%%%%%%%%%%%%%%%%%%%%%%%%%%%%%%%%%%%%%%%%%%%%%%%%%%%%%%%
%                           BIBLIOGRAPHY                               %
%%%%%%%%%%%%%%%%%%%%%%%%%%%%%%%%%%%%%%%%%%%%%%%%%%%%%%%%%%%%%%%%%%%%%%%%

\end{document}